\documentstyle[preprint,eqsecnum,aps]{revtex}

\def\footnoterule{\kern-3pt \hrule width \hsize \kern6.2pt}
\def\tr{~{\rm Tr}~}

\def\pmb#1{\setbox0=\hbox{$#1$}%
\kern-.025em\copy0\kern-\wd0%
\kern.05em\copy0\kern-\wd0%
\kern-.025em\raise.0433em\box0}%

\def\be{\begin{equation}}
\def\ee{\end{equation}}
\def\beq{\begin{eqnarray}}
\def\eeq{\end{eqnarray}}
\def\m{\mu}
\def\n{\nu}
\def\r{\rho}

\def\z{\zeta}
\def\p{\psi}
\def\G{\Gamma}
\def\pa{\partial}
\def\d{\delta}
\def\g{\gamma}
\def\gg{g}

\def\Gc{\Gamma_{\rm c}}
\def\mn{{\m\n}}

\def\abc{{abc}}

\def\dqs{d^{4}\!s\,}
\def\dqk{d^{4}\!k\,}
\def\ln{{\rm ln}\,}
\def\D{{\cal D}}
\def\hq{{\hat q}}
\def\bk{{\bf k}}
\def\disp{\displaystyle}
\def\al{\alpha}
\def\bet{\beta}

\begin{document}

\renewcommand{\thefootnote}{\fnsymbol{footnote}}
\thispagestyle{empty}
\setcounter{page}{0}

\title{HARD THERMAL LOOPS, STATIC RESPONSE AND\\THE COMPOSITE EFFECTIVE
ACTION\footnotemark[1]}

\footnotetext[1]{This work is supported in part by funds
provided by the U.S.~Department of Energy (D.O.E.) under contract
\#DE-AC02-76ER03069.}

\author{R.~Jackiw,~Q.~Liu,~and~C.~Lucchesi\footnotemark[2]}
\footnotetext[2]{Supported by the Swiss National Science
Foundation.}

\address{Center for Theoretical Physics,
         Laboratory for Nuclear Science,
         and Department of Physics\\
         Massachusetts Institute of Technology, Cambridge,
Massachusetts 02139}

\maketitle

\renewcommand{\thefootnote}{\arabic{footnote}}
\setcounter{footnote}{0}
\setcounter{page}{0}
\thispagestyle{empty}

\begin{abstract}
$\!\!$First, we investigate the static non-Abelian Kubo equation.
We prove that it does not possess finite energy solutions; thereby we
establish that gauge theories do not support hard thermal solitons.
This general result is verified by a numerical solution of the equations.
A similar argument shows that ``static" instantons are absent. In addition,
we note that the static equations reproduce the expected screening of the
non-Abelian electric field by a gauge invariant Debye mass
$m=\gg T\, \sqrt{{{N+N_F/2}\over 3}}$. Second, we derive the non-Abelian Kubo
equation from the composite effective action. This is achieved by showing
that the requirement of stationarity of the composite effective action
is equivalent, within a kinematical approximation scheme, to the
condition of gauge invariance for the generating functional of hard thermal
loops.
\end{abstract}

\vfill
\centerline{Submitted to: {\em Physical Review D}}
\vfill
\hbox to \hsize{CTP\#2261 \hfil November 1993}
\vskip-12pt
\eject

\section{Introduction}
\label{sec:1}

When it was realized \cite{ref1} that the gauge invariance condition
\cite{ref2} on the generating functional $\Gamma(A)$ for hard thermal loops in
a gauge theory \cite{ref3} (with or without fermions) coincides with a similar
requirement on the wave functional of Chern-Simons theory, one could use the
known solution for the latter, non-thermal problem \cite{ref4} to give a
construction of $\Gamma(A)$ relevant in the former, thermal context.
The expression for $\Gamma(A)$ is non-local and not very explicit: $\Gamma(A)$
can be presented either as a power series in the gauge field $A$ \cite{ref1}
[the $O(A^n)$ contribution determines the hard thermal gauge field
(and fermion) loop with
$n$ external gauge field lines] or as an explicit
functional of path ordered variables
$P \exp \int d x^\mu \, A_\mu$~ \cite{ref4}.

More accessible is the expression for the induced current
$- {\delta \Gamma(A) \over \delta A_\mu} \equiv
- T^a {\delta \Gamma(A) \over \delta A_\mu^a}$,
which enters (high-temperature) response theory, in a non-Abelian
generalization of Kubo's formula (in Minkowski space-time)
\cite{ref5}:
\be
D_\n \, F^{\n\m}(x)
= - {\delta \Gamma(A) \over \delta A_\m(x)}
\equiv {m^2\over 2} \, j^\m (x)\ .
\label{1.1}
\ee
$T^a$ is an anti-hermitian representation of the Lie algebra,
the gauge covariant derivative is defined as $D_\n =\pa_\n + \gg [A_\n,\ ]$,
and $m$ is the Debye mass determined by the matter
content: in an
$SU(N)$ gauge theory at temperature $T$, with fermions in the
representation
${\cal T}^a$, and $\tr ({\cal T}^a {\cal T}^b) = -{N_F \over 2}
\delta^{ab}$ where $N_F$ counts the number of flavors, the Debye mass
satisfies
\be
m^2 = {\gg^2 T^2\over 3} \, \left( N + {N_F \over 2} \right)\ .
\label{1.2}
\ee
Henceforth, through Section \ref{sec:2}, we scale the gauge coupling constant
to unity.
The functional form of $j^\mu$ can be given as \cite{ref5}
\be
j^\mu(x) = \int {d \hat{q} \over 4\pi}
\, \Biggl\{ Q^\mu_{+} \biggl( a_{-}(x) - A_{-}(x) \biggr) +
Q^\mu_{-} \biggl( a_{+}(x) - A_{+}(x) \biggr) \Biggr\}\ .
\label{1.3}
\ee
Here $Q^\mu_{\pm}$ are the light-like 4-vectors
${1\over\sqrt{2}} (1, \pm \hat{q})$, with $\hat{q}^2=1$,
$A_{\pm}$ are the light-like projections $A_\pm = Q^\mu_{\pm} \, A_\mu$,
while $a_\pm$ are given by \cite{ref4,ref5}
\be
a_+ = g^{-1} \, \partial_{+} \, g ~,~~~~~
a_- = h^{-1} \, \partial_{-} \, h ~~~~~~~
(\partial_\pm \equiv Q^\mu_\pm \, \partial_\mu)
\label{1.5}
\ee
when $A_\pm$ are parametrized as
\be
A_+ = h^{-1} \, \partial_+ \, h ~,~~~~~~
A_- = g^{-1} \, \partial_- \, g ~~.
\label{1.6}
\ee
In other words, $a_\pm$ satisfy the equations
\begin{mathletters}
\label{1.7}
\beq
\partial_+ a_- - \partial_- A_+ + [A_+, a_-] &=& 0\ , \\
\partial_+ A_- - \partial_- a_+ + [a_+, A_-] &=& 0\ ,
\eeq
\end{mathletters}
$\!\!$whose solution can be presented as in (\ref{1.5}) when $A_\pm$ are
parameterized as in (\ref{1.6}) ---
evidently $g$ and $h$ involve path ordered exponential integrals of
$A_\pm$.
(Alternatively $a_\pm$ may be given by a power series in $A_\mp$
\cite{ref1}.)
Finally (\ref{1.3}) requires averaging over the directions of
$\hat{q}$.

It is easy to verify that (\ref{1.7}) ensure covariant conservation of
$j^\mu$.
Moreover, gauge invariance is maintained: for (\ref{1.1}) to be gauge
covariant, it is necessary that $j^\mu$ transform gauge covariantly.
That the
expression in (\ref{1.3}) possesses this property is seen as follows.
When
$A_\pm$ transform by $U^{-1} \, A_\pm \, U + U^{-1} \partial_\pm
U$,
Eqs.~(\ref{1.5}) -- (\ref{1.7}) show that $a_\pm$ transform similarly,
hence the
differences $a_\pm - A_\pm$ transform covariantly. The manifest
gauge
covariance of (\ref{1.1}) ensures that $m$ is a gauge invariant
parameter;
that it also has the interpretation of an electric (Debye) mass will be
evident when we consider the static limit.

It is of obvious interest to discuss solutions of (\ref{1.1}).  In the
Abelian, electrodynamical case this is easy to do, since (\ref{1.7})
can be
readily solved for $a_\pm$, and the solutions of the linear problem
are the
well-known plasma waves \cite{ref6}.  The non-linear
problem of
finding non-Abelian plasma waves is much more formidable.  Also,
one inquires
whether the non-linear equations support soliton solutions, and
(after an
appropriate continuation to imaginary time) instanton solutions.
[The
time-dependent equation (\ref{1.1}) in Minkowski space-time must
be
supplemented with boundary conditions, which are determined by the
physical
context.  For example, response theory requires retarded boundary
conditions,
which in fact preclude deriving (\ref{1.1}) variationally \cite{ref5}.
Here we shall not be concerned with this issue.]

Our paper concerns the following two topics.  In Section~\ref{sec:2}, we
analyze
(\ref{1.1}) for static fields.  It turns out that in the time-independent
case
(\ref{1.7}) can be solved for $a_{\pm}$ and (\ref{1.1}) is presented
in closed form. We prove that the resulting equation does not possess
finite-energy solutions, thereby
establishing that gauge theories do not support hard thermal solitons.
Also some negative conclusions
about instantons are given. In Section~\ref{sec:3} we present an
alternative
derivation of (\ref{1.1}), which relies on the composite effective
action
\cite{ref7}, and makes use of approximations recently developed in an
analysis of hard thermal loops based on the Schwinger-Dyson
equations \cite{ref8}. The Appendix presents a numerical analysis of the
solutions to equation (\ref{1.1}) for $SU(2)$, which supports the
general result in Section~\ref{sec:2}.

\section{Static Response}
\label{sec:2}

When $A_\pm$ are time-independent, we seek solutions of
(\ref{1.7}) that are
also time-independent.  Acting on static fields, the derivatives
$\partial_\pm$
become $\pm {1\over\sqrt{2}} \hat{q} \cdot {\bf \nabla} \equiv
\pm\partial_\tau$,
and (\ref{1.7}) may be written as the equations
\be
\partial_\tau \, {\cal A}_\pm \pm [A_\pm, {\cal A}_\pm] = 0
\label{2.2}
\ee
for the unknowns ${\cal A}_\pm \equiv A_\pm + a_\mp$.
These are solved trivially by ${\cal A}_\pm = 0$, that is
\be
a_\mp = -A_\pm\ .
\label{2.3}
\ee
This solution is also the one that is deduced from the perturbative
series
expression for $a_\pm$, when restricted to static $A_\pm$.

[A non trivial solution can be constructed with the help of
representations
similar to (\ref{1.6}).  Upon defining in the static case
\beq
A_+ &=& h_0^{-1} \, \partial_\tau \, h_0 \ ,\nonumber\\
A_- &=& -g_0^{-1} \, \partial_\tau \, g_0 \nonumber
\eeq
($h_0$ and $g_0$ involve path-ordered exponentials along the path
${\bf r} + \hat{q}\tau$), we find
\beq
{\cal A}_+ = h_0^{-1} \, I_+ \, h_0 \ ,\nonumber\\
{\cal A}_- = g_0^{-1} \, I_- \, g_0 \ ,\nonumber
\eeq
where $I_{\pm}$ are arbitrary Lie algebra elements, independent of
$\tau$: $\hat{q} \cdot {\bf \nabla} I_\pm = 0$.
Since these solutions involve the arbitrary quantities $I_\pm$, and
since they
do not arise in the perturbative series, we do not consider them
further
and remain with the trivial solution (\ref{2.3}), which corresponds to
$I_\pm =0$.]

{}From (\ref{2.3}) it follows that the current for static fields is
\beq
j^\mu({\bf r})
&=& - \int {d\hat{q} \over 4\pi} \,
\biggl( Q^\mu_{+} + Q^\mu_{-} \biggr) \,
\biggl( A_+({\bf r}) + A_-({\bf r}) \biggr) \nonumber\\
&=& - \int {d\hat{q} \over 4\pi} \,
\biggl( Q^\mu_{+} + Q^\mu_{-} \biggr)
\biggl( Q^\nu_{+} + Q^\nu_{-} \biggr)
A_\nu({\bf r})\ .
\label{2.4}
\eeq
With ${\bf Q}_{+} + {\bf Q}_- = 0$ and $Q_{+}^0 + Q_{-}^0 = \sqrt{2}$,
we compute $j^\mu = - 2 \, \delta^{\mu 0}A_0 $.
The response equations (\ref{1.1}) then become, in the static limit:
\begin{mathletters}
\label{2.6}
\beq
{D}_i E^i + m^2 A_0 &=& 0 \ ,\label{2.6a} \\
\epsilon^{ijk} {D}_j B^k &=& [A_0, E^i]\ , \label{2.6b}
\eeq
\end{mathletters}
\noindent
$\!\!$where $E^i \equiv F^{i 0}$ and $F^{ij} \equiv -\epsilon^{ijk} B^k$.
Eqs. (\ref{2.6}) give clear indication that $m$ plays the role of a gauge
invariant,
electric mass.  The fact that the static current is linear in the vector
potential implies the vanishing of hard thermal loops with more that
two
external gauge-field lines, and zero energy --- a fact which can be
checked
from the relevant graphs.

Unfortunately, Eqs.~(\ref{2.6}) do not possess any finite energy
solutions.
This is established by a variant of the argument relevant to the $m^2
= 0$
case \cite{ref9}.

Consider the symmetric tensor
\be
\theta^{ij} = 2 \tr \left( E^i E^j + B^i B^j - {\delta^{ij} \over 2}
(E^2 + B^2 + m^2 A_0^2) \right)\ .
\label{2.7}
\ee
Using (\ref{2.6}) one verifies that for static fields
$\partial_j \, \theta^{ji} = 0$.  Therefore
\be
\int d^3 r \, \theta^{i i} =  \int d^3 r \, \partial_j (x^i \, \theta^{ji})
=  \int dS^j x^i \theta^{ji}\ .
\label{2.8}
\ee
Moreover, the energy of a massive gauge field (with no mass for the
spatial components) can be written as
\begin{mathletters}
\label{2.9}
\be
{\cal E} = \int d^3 r \left\{ - \tr
\left( E^2 + B^2  + {1 \over m^2} (D_i E^i)^2 \right)
+ \tr \left( m A_0 + {D_i E^i \over m} \right)^2 \right\}\ .
\label{2.9a}
\ee
The second trace in the integrand enforces the constraint (\ref{2.6a}).
Consequently, on the constrained surface the energy is a sum of
positive terms \cite{ref10}:
\be
{\cal E} = \int d^3 r
           \left\{ - \tr \left( E^2 + B^2 + m^2 A_0^2 \right) \right\}
\label{2.9b}
\ee
\end{mathletters}
$\!\!$and ${\bf E}$, ${\bf B}$ and $A_0$ must decrease at large distances
sufficiently rapidly so that each of them is square integrable. This in turn
ensures that the surface integral at infinity in (\ref{2.8}) vanishes, so that
static solutions require
\begin{mathletters}
\label{2.10}
\be
\int d^3 r \theta^{i i} = 0\ .
\label{2.10a}
\ee
On the other hand, from (\ref{2.7}), we see that $\theta^{ii}$ is a sum
of positive terms
\be
\theta^{ii} = -\tr ( E^2 + B^2 + 3 m^2 A_0^2 ) \ ,
\label{2.10b}
\ee
\end{mathletters}
$\!\!$hence (\ref{2.10}) imply the vanishing of
$\bf E$, $\bf B$ and $A_0$.

The absence of finite energy static solutions can also be understood
from the
differential equations (\ref{2.6}).  Eq.~(\ref{2.6a}) possesses solutions
for $A_0$ that are either exponentially increasing or decreasing at infinity.
Rejecting
the former removes the freedom of imposing further conditions at
the origin,
and necessarily the exponentially damped solution devolves into one
that is
singular (not integrable) at the origin; see  the Appendix.
(This situation can be contrasted
with, {\it e.g.}, the magnetic dyon solution \cite{ref11}, where
absence of
the mass term allows solutions for $A_0$ with unconstrained large-$r$
behavior, leaving the freedom to select the solution that is regular at the
origin.)

A similar argument shows that there are no ``static'' instanton
solutions.
These would be solutions for which $t$ is replaced by $- i x_4$, $A_0$
by $i
A_4$ and presumably one would seek solutions periodic in $x_4$
with period
$\beta ={1\over T}={1\over m}\sqrt{{{N+N_F/2}\over 3}}\,$.
An $x_4$-independent solution is necessarily
periodic; it would satisfy (\ref{2.6}) with $A_4$ replacing $A_0$
and
opposite sign in the right side of (\ref{2.6b}).
But analysis similar to the above shows that finite-action solutions
do not exist.

\section{HARD THERMAL LOOPS FROM THE COMPOSITE EFFECTIVE
ACTION}
\label{sec:3}

In this Section, we present a derivation of the non-Abelian Kubo
equation (\ref{1.1}) based on the composite effective action of
\cite{ref7}, a generalization of the usual effective action
(obtained by coupling local sources to the fields) in
which one additionally introduces bilocal sources. In the
QCD case, the composite effective action is given by
$S(A)+\Gc(A,G_{\phi})$, where $G_{\phi}(x,y)$ are (undetermined) two-point
functions, and the labels $\phi=A,\p,\z$ denote either gluons, or
fermions-antifermions, or
ghosts-antighosts, respectively (in the end, ghosts play no dynamical role,
beyond maintaining gauge covariance of the final result). $S(A)$ is the pure
Yang-Mills action, and
\beq
\Gc(A,G_{\phi})&=& {{i}\over{2}}
\biggl(\tr\ln G_A^{-1} + \tr\D_A^{-1}G_A\biggr)\nonumber\\
&&-i\biggl(\tr\ln G_\p^{-1}+\tr\D_\p^{-1}G_\p+\tr\ln G_\z^{-1}
+\tr\D_\z^{-1}G_\z\biggr)
\label{3.1}
\eeq
when 2PI contributions are omitted.
The trace is over
space-time arguments as well as over Lorentz and group indices.
The gauge coupling constant $\gg$, which was previously scaled to
unity, is here reinserted.
$\D_\phi^{-1}$ is computed from the usual QCD action $S_{QCD}$
({\it e.g.} in the Feynman gauge):
\be
i\D_{\phi}^{-1}(x,y)=
{\disp{\d^2 S_{QCD}}\over\disp{\d \phi (x)\,\d \phi (y)}}
\label{3.2}
\ee
The fields carry group and space-time
indices, which are symbolically subsumed into the space-time labels
$x,y$.

The truncated composite effective action (\ref{3.1})
comprises the first, dynamical approximation that we make and reflects
the known fact \cite{ref3} that hard thermal loops arise from one-loop graphs.
The full composite effective action of course coincides
with the ordinary effective
action when the two-point functions are evaluated by imposing stationarity
requirements, and the above truncation reproduces the standard
one-loop action involving $\tr\ln\D_\phi^{-1}$.
Nevertheless subsequent analysis is more transparently organized in the
composite effective action formalism.

As indicated in \cite{ref7}, $S+\Gc$ is stationary for physical
processes. This yields the conditions
\begin{mathletters}
\label{3.4}
\beq
D_\n F^{\n\m}  &=& J^{\m}\ , \label{3.4a} \\
G_\phi^{-1} &=& \D_\phi^{-1}\ ,\ \ \ \phi=A,\p,\z \ .\label{3.4b}
\eeq
\end{mathletters}
$\!\!$Computing the local induced current
$J^{\m}(x)=-{{\d\Gc}\over{\d A_\m(x)}}$
involves differentiating $\D_\phi^{-1}$ with respect to $A_\m$. Since
the $\D_\phi^{-1}$ depend locally on $A$, the resulting current is
the local limit of a bilocal expression constructed from the two-point
functions $G_\phi(x,y)$:
\be
J^\m (x)= \lim_{y\rightarrow x} J^\m (x,y)\ ,
\label{3.41}
\ee
where the bilocal current $J^\m(x,y)=T^a J^{\m}_a(x,y)$ is given by
\beq
J^{\m}(x,y)=g\left(\Gamma^{\m}_{\al\bet\g}\,D^\al_x
{\bf G}_{A}^{\ \,\bet\g} (x,y)
+ \pa^\m_y {\bf G}_{\zeta}(x,y)\right)
+ ig T^a \,{\rm tr}\, \gamma^\m{\cal T}^a G_{\psi}(x,y)
\label{3.411}
\eeq
with $\Gamma^\m_{\al\bet\g}\equiv 2g^\m_\bet g_{\al\g} - g^\m_\al g_{\bet\g}
- g^\m_\g g_{\bet\al}$. The trace ``${\rm tr}$" is taken over Dirac spinor
as well as internal symmetry indices, and we have defined
${\bf G}_{A,\zeta}(x,y)=[T^a,T^b] G_{A,\zeta\, ab}(x,y)$ with
$D_x{\bf G}_{A}(x,y)
=\pa_x{\bf G}_{A}(x,y)+g[[A(x),T^b],T^c]G_{A\,bc}(x,y)$.

We now use eqs. (\ref{3.4}) -- (\ref{3.411})
to study ``soft" plasma excitations. ``Soft'' means
that both the energy and the momentum carried by a particle are of
order
$\gg T$, for a coupling constant $\gg\ll 1$, while particles with energy or
momentum
of order $T$ are called hard (see {\it e.g.} \cite{ref3}).
The strategy is to solve the system of
coupled equations (\ref{3.4}), in order to derive from (\ref{3.411})
the expression (\ref{1.3}) for the local current $J^{\m}$.
We approximate eqs. (\ref{3.4}) by expanding them in powers of $\gg$.
The approximation scheme we use was first proposed in \cite{ref8} for
deriving hard thermal loops from the Schwinger-Dyson equations. It
represents an essential step in that derivation. Earlier work on the QCD
plasma (in which this approximation was not used) is reviewed in
\cite{elzeheinz}.
Following \cite{ref8}, we introduce
relative and center of mass coordinates, $s=x-y$ and $X={1\over
2}(x+y)$,
respectively. In these new variables the partial derivatives
carry different dependences on $\gg$: $\pa_s\sim T$ and $\pa_X \sim
\gg T$.
This comes from the fact that $\pa_s$ corresponds to hard loop
momenta,
whereas $\pa_X$ is related to soft external momenta. See \cite{ref8} for a
detailed account.

Next, motivated by the expression (\ref{3.41})--(\ref{3.411}) for the current,
we expand $G_\phi$ in powers of $\gg$:
\be
G_\phi=G_\phi^{(0)}+\gg G_\phi^{(1)}+\gg^2 G_\phi^{(2)}+ ...\ ,
\label{3.42}
\ee
where $G_\phi^{(0)}$ is just the free propagator at temperature $T$ and
$G_\phi^{(i)},\ i\geq 1$ are determined by (\ref{3.4b}).
At leading order in $\gg$ (to which we restrict ourselves in the sequel),
the bilocal current (\ref{3.411})
depends on $G_\phi^{(0)}$  and $G_\phi^{(1)}$:
\beq
J^{\m}_a(X,s) &=& g^2 f^\abc\Biggl[
\Gamma^{\m}_{\al\bet\g}\biggl(\pa^\al_s G_{A\,bc}^{(1) \,\bet\g}(X,s)
+f^{bde}A^\al_d(X)G_{A\,ec}^{(0)\,\bet\g}(X,s)\biggr)
- \pa^\m_s G_{\zeta\,bc}^{(1)}(X,s)\Biggr] \nonumber\\
&&+ ig^2\,{\rm tr}\,\gamma^\m{\cal T}^a G_{\psi}^{(1)}(X,s)
+ \delta J_a^\m(X,s)\ ,
\label{3.43}
\eeq
where $G_\phi(X,s)\equiv G_\phi(X+{s\over 2},X-{s\over 2})$ [and similarly for
$J(X,s)$]. We have added the term $\delta J^{\m}_a(X,s)$ in order to
compensate for the loss of gauge covariance due to non-locality:
\beq
\delta J^{\m}_a(X,s)= g^2 s\!\cdot\! A^b(X)\Biggl[ f^{ace}f^{bcd}
\biggl(3\,\pa_\n^s G_{A\,de}^{(0)\,\m\n}(s)
+ \pa^\m_s G_{\zeta}^{(0)\,de}(s)\biggr)
+i \,{\rm tr}\, {\cal T}^b{\cal T}^a \gamma^\m G_{\psi}^{(0)}(s)\Biggr] .
\label{3.44}
\eeq
Note that this term vanishes in the local limit.

Now, we derive from (\ref{3.4b}) a condition on $G_\phi^{(1)}$.
[It turns out to be convenient to expand,
instead of (\ref{3.4b}), the equivalent equations $\D^{-1}_\phi G_\phi=
G_\phi\D^{-1}_\phi=I$, in which we disregard temperature-independent
contributions.] The ${\cal O}(\gg)$-condition does not fix $G_\phi^{(1)}$
uniquely; hence we need to go to ${\cal O}(\gg^2)$. The condition
so obtained on $G_\phi^{(1)}$ can be used to derive a constraint on the
bilocal current.
The subsequent derivation of this constraint [eq. (\ref{3.10})] is
similar to the one given in \cite{ref8}, to which we refer the
reader for details. Momentum space is most convenient, {\it i.e.}
\be
G_\phi (X,k)=\int \dqs e^{ik\cdot s}G_\phi (X,s)\ ,
\label{3.5}
\ee
the explicit forms for the thermal parts of the free propagators
being ({\it e.g.} in Feynman gauge):
\begin{mathletters}
\label{3.501}
\beq
G_{A\,ab}^{(0)\,\m\n}(k)&=&-2\pi\,\delta^{ab}g^\mn\delta(k^2)\, n_B(k_0)\ ,\\
G_{\psi}^{(0)\,mn}(k)&=&-2\pi\,\delta^{mn} k\!\!\! / \delta(k^2)\, n_F(k_0)\
,\\
G_{\zeta}^{(0)\,ab}(k)&=&\phantom{-}2\pi\,\delta^{ab}\delta(k^2)\, n_B(k_0)\ ,
\eeq
\end{mathletters}
$\!\!$where $n_{B,F}(k_0)=1/(e^{\beta |k_0|} \mp 1)$ are the bosonic
and fermionic probability distributions.

Similarly, for the bilocal current in momentum space one writes
\be
J^\m (X,k)=\int \dqs e^{ik\cdot s}J^\m (X,s)\ .
\label{3.51}
\ee
In the limit $s\rightarrow 0$, or equivalently $y\rightarrow x$,
where $X=x$,
\be
J^\m (x)=J^\m (X) = \int {{\dqk}\over{(2\pi)^4}}\,J^{\m}(X,k)\ .
\label{3.6}
\ee

The resulting constraint on the bilocal current is \cite{ref8}:
\be
Q\cdot D_X\, {J}^{\m}(X,k)=4\pi \gg^2 Q^\m Q^\r k_0 \, F_{\rho 0} \,
\delta(k^2){d\over{dk_0}}[N\,n_B(k_0)+N_F\,n_F(k_0)]\ ,
\label{3.10}
\ee
where $Q^\m\equiv {{k^\m}\over{k_0}} = (1,{\bf Q})$.

Our next task is to make contact between (\ref{3.10}) and the gauge
invariance condition for the generating functional of hard thermal loops.
Our strategy consists in transforming (\ref{3.10}) into two distinct
conditions for positive and negative $k_0$'s, and in taking advantage of
the symmetry properties that arise.
We first integrate the equation (\ref{3.10})
over $|\bk|$ and $k_0\geq 0$. Due to the $\d (k^2)$ on
the right side, the bilocal current is non-vanishing only when
$k_0=|{\bk}|$; hence $\bf Q$ can be replaced by a unit vector
$\hat q\equiv {\bk\over |\bk|}$. The integration thus yields:
\be
Q_+\cdot D_X \, {\cal J}^{\m}_+(X,{\hat q})=- 2 \sqrt{2} \, \pi^3
 m^2 Q_+^\m Q_+^\r F_{\r 0}\ ,\label{3.103}
\ee
where we have defined
\be
{\cal J}^\m_+ (X,\hq) = \int|{\bf k}|^2 d|{\bf k}|\int_0^\infty dk_0\,
J^\m(X,k)\ .
\label{3.1031}
\ee
Similarly, upon introducing
\be
{\cal J}^\m_- (X,\hq) = \int|{\bf k}|^2 d|{\bf k}|\int_{-\infty}^0 dk_0\,
J^\m(X,k)\ ,\label{3.1033}
\ee
the integration of (\ref{3.10}) over $|\bk|$ and $k_0\leq 0$ gives:
\be
Q_-\cdot D_X \, {\cal J}^{\m}_-(X,{\hat q}) = -2 \sqrt{2} \, \pi^3
m^2 Q_-^\m Q_-^\r F_{\r 0}\ ,\label{3.104}
\ee
wherefrom one sees that ${\cal J}^\m_-(X,-\hq)$ satisfies the same
equation (\ref{3.103}) as ${\cal J}^\m_+(X,\hq)$.

Now, using $\int\dqk = \int d\Omega |{\bf k}|^2 d|{\bf k}|dk_0$, we
rewrite the expression (\ref{3.6}) for the local current as
$J^\m(X)=
\int{d{\hat q}\over (2\pi)^4}[{\cal J}^\m_+(X,\hq)+{\cal J}^\m_-(X,\hq)]$.
Here, $\hq$ can be replaced by $-\hq$ in each term of the integrand
separately, since $\hq$ spans the whole solid angle. Therefore, we can write
\be
J^\m (X)=\int{d{\hat q}\over (2\pi)^4}\, {\cal J}^\m(X,\hq)\ ,
\label{3.1041}
\ee
where ${\cal J}^\m(X,\hq)$ is defined as
\be
{\cal J}^\m(X,\hq)\equiv {\cal J}^\m_+(X,\hq) + {\cal J}^\m_-(X,-\hq)\ ,
\label{3.1042}
\ee
and satisfies, as a consequence of (\ref{3.103}) and (\ref{3.104}),
\be
Q_+\cdot D_X \, {\cal J}^{\m}(X,{\hat q})=-4 \sqrt{2} \, \pi^3
 m^2 Q_+^\m Q_+^\r F_{\r 0}\ .
\label{3.11}
\ee
{}From this, after decomposing
\be
{\cal J}^{\m}(X,{\hat q})={\tilde {\cal J}}^{\m}(X,{\hat q}) - 4 \sqrt{2}
\, \pi^3 m^2 Q_+^\m A_0\ ,
\label{3.13}
\ee
we get as our final condition on the bilocal current:
\be
Q_+\cdot D_X \, {\tilde {\cal J}}^{\m}(X,{\hat q})=4 \sqrt{2} \, \pi^3
m^2 Q_+^\m \pa^0_X (Q_+\cdot A)\ .
\label{3.14}
\ee

Let us now assume that
${\tilde {\cal J}}^{\m}(X,\hq)$ can be obtained from a functional
$W(A,\hq)$ as
\be
{\tilde {\cal J}}^{\m}(X,\hq)={{\d W(A,\hq)}\over{\d A_\m(X)}}\ .
\label{3.15}
\ee
Equation (\ref{3.14}) then implies that $W(A,\hq)$ depends only on
$A_+$, {\it i.e.} $W(A,\hq) =W(A_+)$,
and ${\tilde {\cal J}}^{\m}={{\d W( A_+ )}\over{\d A_+ }}Q_+^\m$.
In turn, $W( A_+ )$ satisfies, as a consequence of (\ref{3.14}),
\be
Q_+ \cdot D_X \, {{\d W( A_+ )}\over{\d  A_+ }}= 4 \sqrt{2} \, \pi^3
m^2 \pa^0_X  A_+  \ .
\label{3.16}
\ee
By introducing new coordinates $(x_+,x_-,{\bf x}_\bot )$,
\be
x_+= Q_-\cdot X,\qquad  x_-=Q_+ \cdot X,\qquad
{\bf x}_\bot \cdot {\hat q} =0\ ,
\label{3.17}
\ee
we can rewrite $Q_+ \cdot \pa_X$ as $\pa_+$ and (\ref{3.16})
becomes
\be
\pa_+ \, {{\d W( A_+ )}\over{\d  A_+ }}
+\gg \left[ A_+ \, ,{{\d W( A_+ )}\over{\d  A_+ }}\right]
=4 \sqrt{2} \, \pi^3 m^2 \pa^0_X A_+ \ .
\label{3.18}
\ee
This equation was first derived in \cite{ref2}, as an expression of
gauge invariance of the generating functional for hard
thermal loops, and has since then been studied by several authors.
Here, it is seen to be a consequence of the stationarity requirement on
the composite effective action.

It has been shown in \cite{ref1} that $W( A_+ )$ is given by the
eikonal of a Chern-Simons gauge theory. This observation is our last step
towards deriving the approximate expression for the local current $J^\m(x)$ in
eq. (\ref{3.4a}).
The subsequent development follows \cite{ref5} and the result is exactly the
non-Abelian Kubo equation (\ref{1.1}) with the form (\ref{1.3}) for the induced
current.

\section{CONCLUSIONS}
\label{sec:4}

The behavior of the quark-gluon plasma at high temperature is described by the
non-Abelian Kubo equation (\ref{1.1}) -- (\ref{1.3}). We have
studied the static response of such a plasma and proved that there are no
hard thermal solitons (this result is supported and illustrated by
numerical integration). The absence of ``static" instantons is established by
invoking a similar argument. In addition, the static non-Abelian Kubo
equation indicates that the non-Abelian electric field is screened by a
gauge invariant Debye mass $m=\gg T\, \sqrt{{{N+N_F/2}\over 3}}$.

Furthermore, we have derived  the non-Abelian Kubo equation from the
composite effective action formalism. Indeed, the requirement that the
composite effective action be stationary leads, within a kinematical
approximation scheme taken at the leading order, to the equation
obtained in \cite{ref2} by imposing gauge invariance on the
generating funtional of hard thermal loops.

Let us mention some problems deserving further investigation.
Finding non-static solutions to the non-Abelian Kubo equation is an
appealing --- if difficult --- task, since
such solutions would correspond to collective excitations of the
quark-gluon plasma at high temperature. Also, it would be interesting to
investigate the next-to-leading order effects in the kinematical
approximation and to see if they are gauge invariant; we hope that
our formalism is well suited for such an investigation. Furthermore,
it is clear that $\Gc(A,G_{\phi})$, when evaluated on the solution for
$G_{\phi}$ obtained from (\ref{3.4b}) and (\ref{3.42}), coincides with the
$\G(A)$ constructed from the Chern-Simons eikonal. While our derivation
establishes this
fact indirectly, an explicit evaluation of the relevant functional
determinants in the hard thermal limit would be welcome.

\vskip0.8truecm
\centerline{\bf NOTE ADDED}
\vskip0.7truecm
\setcounter{equation}{0}
\makeatletter
\def\theequation@prefix{N}
\makeatother

We have now seen recent papers \cite{BInew}
wherein the response equations
are also analyzed.  Moreover, local equations are found
for time-dependent, but space-independent gauge
fields, and for non-Abelian plane waves.
The starting point of these investigations is a non-local
expression for the induced current (see \cite{ref8,BInew}),
\be
j^{\rm ind}_{\mu\ a}(x)=3\,\omega_p^2\int
{{d\Omega}\over{4\pi}}\,v_\mu \int_0^\infty du\,U_{ab}(x,x-vu)
\,{\bf v}\cdot{\bf E}^b(x-vu)\ ,
\label{A.1}
\ee
which appears different from our local, but coupled, form
(\ref{1.3}) -- (\ref{1.7}). Here we exhibit the steps that
explicitly relate the two.

Beginning with our form for the induced current,
(\ref{1.3}) -- (\ref{1.7}), we observe that, owing to
the integration over the angles of $\hat q$, we may collapse these
expressions into
\be
{m^2\over 2}\,j^{\mu}(x) = m^2\,\int {d \hat{q} \over 4\pi}
\, Q^\mu_{+} \biggl( a_{-}(x) - A_{-}(x) \biggr) \ ,
\label{A.2}
\ee
where
\be
\partial_+ a_- + [A_+,a_-] = \partial_- A_+ \ .
\label{A.3}
\ee
Eq. (\ref{A.3}) may be integrated, yielding
\be
a^a_-(x) = \int_0^\infty du\,U_{ab}(x, x - Q_+ u)\,
\partial_-  A_+^b(x - Q_+ u) \ .
\label{A.4}
\ee
Here $U_{ab}$ satisfies
\beq
{\partial\over \partial u} U_{ab}(x,x-Q_+u)
&=&U_{ac}(x,x-Q_+u)\,f_{cbd}\,A_+^d(x-Q_+u)\ ,\nonumber\\
U_{ab}(x,x)&=&\delta_{ab}\ .
\eeq
Also $A_-^a$ may be presented as
\beq
A^a_-(x) &=& - \int_0^\infty du \, {d\over du} \biggl\{
U_{ab}(x, x-Q_+ u)\, A^b_-(x-Q_+ u)\biggr\}\nonumber\\
&=& \int_0^\infty du \, U_{ab}(x, x-Q_+ u)
\biggl\{ \partial_+ A_-^b(x-Q_+u)\nonumber\\
&&\phantom{\int_0^\infty}
- f^{bcd}\, A_-^c(x-Q_+u)\,A_+^d(x-Q_+u)\biggr\}.
\label{A.5}
\eeq
[We have assumed that no contributions
arise at infinity.] From (\ref{A.2}), (\ref{A.4}) and (\ref{A.5}),
it follows that the induced current can be written as
\be
{m^2\over 2}\,j^{\mu}_a(x) =  m^2 \int {d{\hat q}\over 4\pi}
\,Q^\mu_+ \int_0^\infty du\,U_{ab}(x, x-Q_+u)\
F_{-+}^b(x-Q_+u)\ ,
\label{A.6}
\ee
which coincides with the expression (\ref{A.1}) derived in \cite{BInew},
after the notational replacements
$m \rightarrow \sqrt{3}\,\omega_p$, $d{\hat q} \rightarrow d\Omega$,
$Q^\mu_+\rightarrow
v^\mu$ and $F_{-+}\rightarrow {\bf v}\cdot{\bf E}$ are performed.

The time-dependent, space-independent equation found in \cite{BInew}
is easily derived in
our formalism, also. When there is no space dependence, eqs. (\ref{1.7})
can be written as
\be
\partial_+(a_\mp - A_\pm)+[A_\pm,a_\mp -A_\pm] = 0
\label{A.7}
\ee
and are solved by $a_\mp = A_\pm$. Hence:
\be
{m^2\over 2}\,j^{\mu} = {{m^2}\over 2}\int {d{\hat q}\over 4\pi}\,
(Q_+ - Q_-)^\mu (Q_+ - Q_-)^\nu A_\nu \ ,
\label{A.8}
\ee
of which only the spatial component is non-vanishing:
\be
{m^2\over 2}\,j^i = {m^2} \int {d{\hat q}\over 4\pi}\,
{\hat q}^i {\hat q}^j A_j
=- {1\over 3}\, m^2 A^i\ .
\label{A.9}
\ee
This coincides with the result in \cite{BInew}.

Similarly, the induced current for the non-Abelian plane wave in
\cite{BInew} corresponds to:
\be
a_\pm = {Q_\pm\cdot p\over Q_\mp\cdot p}\, A_\mp (p\cdot x)
\ee
in our formalism, with $p=(\omega,{\vec k})$ being the corresponding
wave vector.

\vskip0.8truecm
\centerline{\bf APPENDIX}
\vskip0.7truecm
\setcounter{equation}{0}
\makeatletter
\def\theequation@prefix{A}
\makeatother

In this Appendix we analyze in greater detail and integrate numerically the
radially symmetric version of the static response equations (\ref{2.6}), in
the $SU(2)$ case. Radially symmetric $SU(2)$ gauge potentials
take the forms:
\beq
\label{eq:a1}
A^a_i &=& ( \delta^{ai} - \hat{r}^a \hat{r}^i ) \, {\phi_2(r) \over r} +
\varepsilon^{aij} \, \hat{r}^j \, {{1 - \phi_1(r)} \over r}\ , \nonumber \\
A_0^a &=& \hat{r}^a \, { g(r) \over r}\ ,
\eeq
where a residual gauge freedom has been used to eliminate
a term proportional to $ \hat{r}^a \hat{r}^i$.

We substitute the {\it Ansatz} (\ref{eq:a1}) into (\ref{2.6}).
The resulting equations
give us the freedom to set one of the two $\phi_i$'s to zero; we
obtain,
\beq
\label{eq:a3}
x^2 {d^2 \over dx^2}\, J &=& (x^2 + 2 K^2) \, J\ , \nonumber \\
x^2 {d^2 \over dx^2}\, K &=& (K^2 -J^2 -1) \, K\ ,
\eeq
where we have set $\phi_2$ to zero, rescaled $x=mr$ and defined
$J(x)=g(r)$, $K(x)=\phi_1(r)$.

We now investigate this system of coupled second-order differential
equations. First, we see
that they possess the following two exact solutions:
\begin{mathletters}
\beq
J&=&0,~K=\pm 1 \ , \label{eq:a4a} \\
J&=&J_0\,e^{-x},~K=0\ . \label{eq:a4b}
\eeq
\end{mathletters}
$\!\!$Eq.~(\ref{eq:a4a}) corresponds to the Yang-Mills vacuum, while
(\ref{eq:a4b}) is the celebrated Wu-Yang monopole plus a screened electric
field.

In the asymptotic region $x\rightarrow\infty$, the
regular solution of the system (\ref{eq:a3}) tends to (\ref{eq:a4a}), with
$J$ approaching its asymptote exponentially. (Of course there is also the
solution with $J$ growing exponentially, which we do not consider.)

Near the origin, $J$ and $K$ behave either like the vacuum (\ref{eq:a4a}) or
approach the monopole solution (\ref{eq:a4b}) as follows,
\beq
J(x) &\rightarrow& J_0 + ... \ ,\nonumber\\
K(x) &\rightarrow& K_0\sqrt{x} \, cos\biggl( {2\pi\over \tau} ln {x \over x_0}
\biggr)+ ...\ ,
\label{eq:a44}
\eeq
where $\tau$ is correlated with $J_0$ as
\be
\tau={4\pi\over\sqrt{4J_0^2+3}}\ .
\label{eq:a45}
\ee
Only the vacuum alternative at the origin leads to finite energy.
However, since we must choose one of two possible solutions at infinity
(obviously we pick the regular one), the behavior at
the origin is determined and can be exhibited explicitly by integrating the
equations (\ref{eq:a3}) numerically. Starting
with regular boundary conditions at infinity, we find the profiles
presented in Figure 1. They show that the monopole solution (\ref{eq:a4b})
is reached at the origin, with $K$ vanishing as in (\ref{eq:a44})
-- (\ref{eq:a45}),
a result consistent with our analytic proof that there are no
finite energy static solutions in hard thermal gauge theories.

\end{document}